\documentclass[letterpaper,times]{IONconf}
\newcommand{\keyword}[1]{\textbf{Keywords:} #1}
\usepackage{setspace}
\usepackage{url}

\usepackage[hidelinks]{hyperref}

\usepackage{siunitx}
\usepackage{lipsum}
\usepackage{amsmath}
\usepackage{amsfonts}
\usepackage{bm}
\usepackage{tabularray}
\usepackage{graphicx}
\usepackage{graphicx}
\usepackage{float}
\usepackage{subfig}
\usepackage{environ}
\NewEnviron{review}{\textcolor{black}{\BODY}}
\NewEnviron{dummy}{\textcolor[rgb]{0.8,0.8,0.8}{\BODY}}
\usepackage{threeparttable}
\usepackage{csquotes}

\DeclareMathOperator*{\E} {\mathrm{E}}
\DeclareMathOperator*{\Var} {\mathrm{Var}}
\DeclareMathOperator*{\Cov} {\mathrm{Cov}}
\newcommand*\diff{\mathop{}\!\mathrm{d}}

\usepackage[left]{lineno}

\title{An Efficient Detector for Faulty GNSS Measurements Detection With Non-Gaussian Noises}

\author{
    Penggao Yan, Baoshan Song, Xiao Xia, Weisong Wen, and Li-Ta Hsu* \\
    \textit{Department of Aeronautical and Aviation Engineering, The Hong Kong Polytechnic University, Hong Kong} \\
    lt.hsu@polyu.edu.hk
    }

\begin{document}

\maketitle

\pagestyle{myheadings}
\setcounter{page}{1}

\section*{Abstract}
Fault detection is critical for reliable navigation, but most existing methods rely on Gaussian error assumptions and struggle under non-Gaussian conditions. This paper proposes the jackknife detector, a computationally efficient method for fault detection in linearized pseudorange-based positioning systems with non-Gaussian nominal errors. By leveraging the jackknife resampling technique, the detector constructs a test statistic as a linear combination of measurement errors, avoiding restrictive distributional assumptions while enabling real-time implementation. A hypothesis test with Bonferroni correction is developed to identify faulty measurements. Theoretical analysis shows that the jackknife detector is equivalent to the widely used solution separation (SS) detector but is significantly more efficient. Through worldwide simulations, Monte Carlo analysis, and real-world satellite clock anomaly detection—all with non-Gaussian errors—the jackknife detector demonstrates comparable detection performance to the SS detector, with a threefold gain in computational efficiency. These results underscore its promise for real-time, safety-critical applications in non-Gaussian noise environments.

\keyword{Fault detection, Jackknife test, non-Gaussian errors, Global navigation satellite system}

\section{INTRODUCTION}
Global Navigation Satellite Systems (GNSS) serve as the backbone for safety-critical applications ranging from aviation to autonomous vehicles, where positioning failures can have catastrophic consequences. Given these high stakes, robust fault detection becomes paramount to ensure system reliability and prevent potentially disastrous outcomes \citep{osechas2012carrier,pervan_multiple_1998,joerger_fault_2016,gao2023solution,sun2021new,yan2025fault}. This technology encompasses both the detection of fault occurrence and the determination of fault timing \citep{gao_real-time_2015}, which are essential for maintaining operational safety in mission-critical environments. 

For pseudorange-based positioning systems, such as the single point positioning (SPP) and the differential global navigation satellite system (DGNSS) positioning, the fundamental principle of fault detection is to identify measurements that deviate from the system’s expected behavior, where abnormal measurements usually show a considerable inconsistency with normal measurement \citep{hsu2017multiple,kuusniemi2004position}. Based on this principle, fault detection methods such as the range comparison method \citep{lee1986analysis}, parity space \citep{sturza1988navigation,pervan1996parity}, chi-squared test \citep{walter_weighted_1995,joerger2013kalman}, and solution separation (SS) \citep{brown1988self,blanch2010raim},  are developed and widely applied in satellite-based navigation systems for fault detection and integrity monitoring purposes \citep{blanch2015baseline,joerger_solution_2014,li2017integrity,hewitson2006gnss}. However, a common assumption has been made in developing these methods that the nominal measurement error is Gaussian distributed. Indeed, this assumption brings several benefits and charming properties in constructing and applying these fault detection methods, such as 1) uncorrelated Gaussian variables are independent, which is the foundation for determining the theoretical threshold of the parity space \citep{sturza1988navigation} and the chi-squared methods \citep{walter_weighted_1995}; 2) the linear combination of Gaussian variables is still Gaussian distributed, which makes it much easy to project measurement domain errors to position domain errors and thus favorite in receiver autonomous integrity monitoring (RAIM) \citep{blanch2010raim}. However, measurement errors or noises in the real world usually have non-Gaussian properties. For example, the noise of global navigation satellite system (GNSS) signals exhibits a strong non-Gaussian property, as exemplified in many studies \citep{niu_using_2014,rife_core_2004,braff2005method,wu2017long}. This distributional mismatch can result in degraded fault detection rates in real-world applications, limiting the reliability and effectiveness of preventing systems from faults.\par

Despite extensive research addressing non-Gaussian conditions in state estimation through robust estimation \citep{sunderhauf2013switchable,wen_factor_2021,pfeifer2021advancing}, particle filters \citep{gabela2021case,ray2018comparison,gupta2022getting},, and adaptive error modeling \citep{pfeifer2019expectation,yan2024subspace}, fault detection under non-Gaussian nominal errors remains largely underexplored. The early exploration of non-Gaussian fault detection methods can be roughly classified into two categories: the Gaussian sum filter (GSF) approach and the particle filter (PF) approach. In the first category, \cite{yun_gaussian_2008} use the Gaussian mixture model (GMM) to model measurement errors and deploy several parallel Kalman filters to deal with each Gaussian component. The fault detection process is realized by comparing the one-side tail probability of the GMM-distributed residual obtained from the GSF with a threshold, which is heuristically obtained by constructing a Kalman filter based detector under Gaussian assumptions. \cite{wang_protection_2022} also develop a similar algorithm. The difference is that Wang's approach sums up the residual of each Kalman filter according to the mixture weight and subsequently takes the summation for a Chi-squared test. However, an assumption is made yet not proved in \cite{wang_protection_2022} that the weighted summed residual is non-central chi-square distributed. In the second category, a common logic is applied in most of the studies. The measurement set is divided into several subsets within which one measurement is excluded (similar to the concept of SS). Then, for the full measurement set and each subset, a PF is applied to estimate the state and the likelihood of the predicted state. Finally, a cumulative log-likelihood ratio (LLR) test is constructed by using the likelihood produced by each auxiliary PF and the main PF. Differences among PF-based fault detection literature mainly lie in the determination of the threshold. \cite{rosihan2006particle} set the threshold with an empirical value; \cite{he2016nonlinear} develop a heuristic approach based on the genetic algorithm to solve the near-optimal solution for the threshold; \cite{wang2018fault} conduct a simulation to determine the threshold. Notably, the PF-based approach is computationally intensive when the number of particles increases. In summary, these early explorations of non-Gaussian fault detection methods either lack rigorous statistical properties or are computationally expensive, limiting their reliability in safety-critical applications. \par

To address these fundamental limitations, this paper introduces the jackknife detector, a statistically rigorous approach that eliminates distributional assumptions while maintaining computational efficiency. The idea is to formalize a rigorous hypothesis testing under non-Gaussian nominal errors by introducing the jackknife technique, a cross-validation technique in statistics \citep{tukey1958bias, quenouille1956notes}. Specifically, we compute the jackknife residual as the inconsistency between the observed measurement and the predicted measurement based on subset solutions, which is proven to be the linear combination of measurement errors (Section \ref{subsec:jackknifeGaussian}). Using this jackknife residual as the test statistic, a jackknife detector is then developed by formalizing a multiple-testing problem with the Bonferroni correction \citep{bonferroni1936teoria} (Section \ref{sec:test}). We further prove that the jackknife detector and the SS detector are theoretically equivalent, but the jackknife detector is more efficient in computation when the nominal error models are non-Gaussian (Section \ref{sec:relation}). In a worldwide simulation, the proposed jackknife detector demonstrates the equivalent performance with the SS detector in detecting faulty measurements in a dual-frequency SPP system (Section \ref{sec:world simulation}). Additionally, Monte Carlo analysis with 1,000 simulations provides rigorous statistical validation of the theoretical equivalence between jackknife and SS detectors (Section \ref{sec:MC}). While both methods demonstrate comparable detection performance, occasional minor discrepancies (the jackknife detector occasionally shows marginally better detection performance than the SS detector when nominal error models are non-Gaussian) highlight the sensitivity of hypothesis testing to numerical implementation differences in non-Gaussian distribution handling. Moreover, we apply the proposed method in a single-frequency SPP system to detect real-world satellite clock anomalies, where the nominal measurement error shows significantly heavy tails. The results reveal that the jackknife detector substantially reduces the computational load compared to the SS detector, while maintaining comparable performance in anomaly detection (Section \ref{sec:realworld}). However, the proposed method is primarily designed for single-fault scenarios and assumes zero-mean nominal errors, which represent areas for future research. This work extends our conference paper \citep{yan2024jackknife} by theoretically establishing the equivalence between jackknife and solution separation detectors and evaluating the computational efficiency advantages for real-time applications. The contributions of this paper are threefold:
\begin{enumerate}
    \item Proposes the jackknife detector, which provides a computationally efficient approach for detecting faults in linearized pseudorange-based positioning systems under non-Gaussian nominal errors, eliminating the need for restrictive distributional assumptions that plague existing methods;
    \item Establishes the mathematical relationship between the jackknife and SS detectors, revealing that the SS detector is the projection of the jackknife detector along the direction defined by the derivative of the solution with respect to the suspected measurement, while demonstrating the jackknife detector's superior computational efficiency in non-Gaussian environments;
    \item Validates the practical applicability of the proposed method through simulated experiments and demonstrates its real-time capability in detecting real-world satellite clock anomalies, achieving significant improvements in both detection delay and computational performance.
\end{enumerate}

\section{JACKKNIFE DETECTOR FOR LINEAR SYSTEM}\label{sec:jack}
\subsection{Linearization of Pseudorange-based Positioning Systems}\label{subsec:linearSystem}
The GNSS pseudorange measurement model is a non-linear system, which can be formalized as follows \citep{misra_global_2006}:
\begin{equation}
    \varrho_i = \sqrt{\left(p_{x}^{i}-u_x\right)^2+\left(p_{y}^{i}-u_y\right)^2+\left(p_{z}^{i}-u_z\right)^2} +u_t+\eta_i \,,
\label{eq:meas_model}
\end{equation}
where $\varrho_i$ is the $i$th pseudorange measurement, $\mathbf{p}^{i} = \left[p_{x}^{i},p_{y}^{i},p_{z}^{i}\right]^T$ is the position of $i$th satellite, $u_x$, $u_y$, and $u_z$ are the receiver position in the Earth-Centered, Earth-Fixed (ECEF) coordinate system, $u_t = c\delta_r$, $\delta_r$ is the receiver clock bias from a single satellite constellation, $c=3\times10^8~\si{\meter/\second}$ is the speed of light, and $\eta_i$ is the measurement error. \par

In most GNSS positioning applications, such as SPP and DGNSS positioning, the pseudorange measurement model is linearized by taking the first-order Taylor expansion at a certain linearized point $\mathbf{x}_0$. A generalized linear system for GNSS positioning can be written as
\begin{equation}
\mathbf{y}=\mathbf{G}\mathbf{x}+\bm{\varepsilon} \,,
\label{eq:general_linear}
\end{equation}
where
\begin{equation}
\begin{aligned}
    \mathbf{y} =\begin{bmatrix}
        f\big(\rho_1,\mathbf{x}_0\big)\\
        \vdots \\
        f\big(\rho_n,\mathbf{x}_0\big)
    \end{bmatrix}\,,
    &\mathbf{G} = \begin{bmatrix}
     \mathbf{g}\big(\{p^{1,j}\},\mathbf{x}_0\big)\\
     \vdots  \\
    \mathbf{g}\big(\{p^{n,j}\},\mathbf{x}_0\big)\\
    \end{bmatrix} \,,
    \bm{\varepsilon}=\begin{bmatrix}
        \varepsilon_1 \\
        \vdots \\
        \varepsilon_n
    \end{bmatrix} \,, \\
    &\mathbf{x}{=} \mathbf{x}_t - \mathbf{x}_0\,,
\label{eq:general_linear2}
\end{aligned}
\end{equation}
$\mathbf{x}$ is the system state; $\mathbf{x}_t$ is the receiver positioning state (an $m\times 1$ vector) and $\mathbf{x}_0$ is its linearized point; $\varepsilon_i$ is the $i$th measurement error; $f\big(\rho_i,\mathbf{x}_0\big)$ is a function of the $i$th measurement $\rho_i$ (note that $\rho_i$ refers to a generalized measurement, not limited to the pseudorange measurement) and $\mathbf{x}_0$; and $\mathbf{g}\big(\{p^{i,j}\},\mathbf{x}_0\big)$ is a vector function of the collection of satellite positions $\{p^{i,j}\}$ related to the $i$th measurement and $\mathbf{x}_0$. For the SPP system,  $\{p^{i,j}\}$ involves one satellite; for the DGNSS positioning system,  $\{p^{i,j}\}$ involves two satellites. The exact form of Eq. \eqref{eq:general_linear2} for the SPP system is given in Appendix \ref{app:linearform}. In the following, this paper utilizes this general expression to develop the jackknife detector for detecting faulty measurements in pseudorange-based positioning systems. 

\subsection{Construction of Jackknife Residual}\label{subsec:jackknifeGaussian}
In statistics, the jackknife is a cross-validation technique, initially developed by \cite{quenouille1949problems} and expended and named by \cite{tukey1958bias}. The basic idea of the jackknife technique is to systematically leave out each observation from a dataset and calculate the parameter estimate over the remaining observations. Then, these calculations are aggregated for specific statistical purposes \citep{tukey1958bias, quenouille1956notes}. This section shows how to derive the jackknife residual for linearized pseudorange-based positioning systems and develop the hypothesis test to detect potential faults.\par
\subsubsection{Full set and subset solutions based on weighted least square}
With $n$ measurements, the estimated receiver positioning state $\hat{\mathbf{x}}_t$ can be solved by the weighted least square (WLS) method (in an iterative approach) as follows:
\begin{equation}
\begin{aligned}
    \hat{\mathbf{x}} &=  \mathbf{S} \mathbf{y} \\
    \hat{\mathbf{x}}_t &=  \mathbf{x}_0 + \hat{\mathbf{x}} \,,
\end{aligned}
\label{eq:full solution}
\end{equation}
where $\hat{\mathbf{x}}$ is the full set solution, $\mathbf{S}$ is the weighted least square solution matrix
\begin{equation}
 \mathbf{S}=\left(\mathbf{G}^T\mathbf{W}\mathbf{G}\right)^{-1}\mathbf{G}^T\mathbf{W} \,,
\end{equation}
and $\mathbf{W}$ is the weighting matrix and usually takes the inverse of the covariance matrix of $\varepsilon$. Define the solution matrix for the $k$th subset as follows:
\begin{equation}
    \mathbf{S}^{(k)} = (\mathbf{G}^T\mathbf{W}^{(k)}\mathbf{G})^{-1}\mathbf{G}^T\mathbf{W}^{(k)} \,,
    \label{eq:S^i}
\end{equation}
where $\mathbf{W}^{(k)}$ is a diagonal matrix and is defined as 
\begin{equation}
    W^{(k)}_{i,i} = \begin{cases}
        0 & \text{if}~ i=k \\
        W_{i,i}  & \text{otherwise}
    \end{cases} \,.
\end{equation}
The subsolutions are then given by
\begin{subequations}
  \begin{align} 
  \hat{\mathbf{x}}^{(k)} {=}&  \mathbf{S}^{(k)} \mathbf{y} ~ \forall k=1\cdots n \\
  \hat{\mathbf{x}}^{(k)}_t {=}& \mathbf{x}_{0}+\hat{\mathbf{x}}^{(k)} ~ \forall k=1\cdots n \,,
  \end{align}
  \label{eq:subsolution}
\end{subequations}
where $ \hat{\mathbf{x}}^{(k)}_t$ is the estimation of the positioning state $\mathbf{x}_t^{(k)}$ associated with the $k$th subset.

\subsubsection{Construction of Jackknife residual}
The predicted $k$th measurement with the subsolution $\hat{\mathbf{x}}^{(k)}$ is given by 
\begin{equation}
\hat{y}_k = \mathbf{g}_k \hat{\mathbf{x}}^{(k)} \,,
\label{eq:prediction}
\end{equation}
where $ \mathbf{g}_k$ is the $k$th row of $\mathbf{G}$. The jackknife residual is given by the difference between $y_k$ and $\hat{y}_k$
\begin{equation}
    t_k=y_k - \hat{y}_k \,,
    \label{eq:JK_residual}
\end{equation}
where $y_k$ is the $k$th element of $\mathbf{y}$. \par

\subsubsection{Distribution of jackknife residual}\label{subsec:ex_JK}
The predicted measurement vector $\tilde{\mathbf{y}}^{(k)}$ based on the subsolution $ \hat{\mathbf{x}}^{(k)}$ is given by
\begin{equation}
    \tilde{\mathbf{y}}^{(k)} = \mathbf{G}\hat{\mathbf{x}}^{(k)} \,.
\end{equation}
Then the measurement residual vector is given by
\begin{equation}
    \begin{aligned}
        \mathbf{y}-\tilde{\mathbf{y}}^{(k)} {=}& \mathbf{y} - \mathbf{G}  \hat{\mathbf{x}}^{(k)} \\
        {=}& \left(\mathbf{I}-\tilde{\mathbf{P}}^{(k)}\right)\mathbf{y} \,,
    \end{aligned}
    \label{eq:res_vec}
\end{equation}
where
\begin{equation}
    \tilde{\mathbf{P}}^{(k)} = \mathbf{G} \mathbf{S}^{(k)} \,.
    \label{eq:P_tilde}
\end{equation}
By substituting Eq. \eqref{eq:general_linear} into Eq. \eqref{eq:res_vec}, we have
\begin{equation}
   \mathbf{y}-\tilde{\mathbf{y}}^{(k)} = \left(\mathbf{I}-\tilde{\mathbf{P}}^{(k)}\right)\mathbf{G} \mathbf{x}^{(k)} + \left(\mathbf{I}-\tilde{\mathbf{P}}^{(k)}\right) \bm{\varepsilon} \,.
\end{equation}
Since
\begin{equation}
\begin{aligned}
    \left(\mathbf{I}-\tilde{\mathbf{P}}^{(k)}\right)\mathbf{G} &= \left(\mathbf{I}-\mathbf{G} \mathbf{S}^{(k)}\right)\mathbf{G} \\
    & = \left(\mathbf{I}-\mathbf{G} (\mathbf{G}^T\mathbf{W}^{(k)}\mathbf{G})^{-1}\mathbf{G}^T\mathbf{W}^{(k)}\right)\mathbf{G} \\ 
    &=0 \,,
\end{aligned}
\end{equation}
we have
\begin{equation}
    \mathbf{y}-\tilde{\mathbf{y}}^{(k)} = \left(\mathbf{I}-\tilde{\mathbf{P}}^{(k)}\right) \bm{\varepsilon} \,.
\end{equation}
Define $\tilde{\mathbf{p}}^{(k)}_k$ as the $k$th row of $\left(\mathbf{I}-\tilde{\mathbf{P}}^{(k)}\right)$, then the jackknife residual is given by 
\begin{equation}
    t_k= \tilde{\mathbf{p}}^{(k)}_k \bm{\varepsilon} \,.
    \label{eq:ti_2}
\end{equation}
Eq. \eqref{eq:ti_2} can be rewritten as the linear combination of measurement errors as follows:
\begin{equation}
    t_k = \sum\limits_{j=1}^{n} \tilde{p}^{(k)}_{k,j}\varepsilon_j \,,
    \label{eq:ti_3}
\end{equation}
where $\tilde{p}^{(k)}_{k,j}$ is the $j$th element of $\tilde{\mathbf{p}}^{(k)}_k$, and $\varepsilon_j$ is the $j$th element of $\bm{\varepsilon}$. Remarkably, $\varepsilon_j$  can have an \textbf{arbitrary distribution} as long as it has a PDF $f_{\varepsilon_j}(\cdot)$. Since $t_k$ is the weighted sum of independent random variables with zero-mean distributions, its PDF can be easily obtained by \citep{lee_jiyun_sigma_2009,yan2024principal}
\begin{equation}
    f_{t_k}(x) = {\prod\limits_{j = 1}^{n}\left| \tilde{p}^{(k)}_{k,j} \right|^{- 1}} f_{\varepsilon_1}\left( \frac{x}{\left| \tilde{p}^{(k)}_{k,1} \right|} \right)*f_{\varepsilon_2}\left( \frac{x}{\left| \tilde{p}^{(k)}_{k,2} \right|} \right)*\ldots  * f_{\varepsilon_n}\left( \frac{x}{\left| \tilde{p}^{(k)}_{k,n} \right|} \right) \,.
    \label{eq:ti_conv}
\end{equation}
where $*$ denotes the convolution operation.

In the special case where  $\varepsilon_j$ has a zero-mean Gaussian distribution, i.e., 
\begin{equation}
    \varepsilon_j \sim \mathcal{N}\left(0,\sigma_j^2\right) ~ \forall j=1\cdots n \,,
    \label{eq:GaussianNoise}
\end{equation}
the distribution of $t_k$ is given by (a proof is provided in Appendix \ref{app:JK_res_proof})
\begin{equation}
    t_k \sim \mathcal{N}\left(0,\mathbf{g}_k \mathbf{S}^{(k)}\mathbf{W}^{-1}\mathbf{S}^{(k)^T} \mathbf{g}_k^T + \sigma_k^2 \right) \,.
    \label{eq:ti_dis}
\end{equation}

\subsection{Jackknife test for fault detection}\label{sec:test}
Formalize the following hypotheses:
\begin{equation}
\begin{aligned}
    H_0^{(k)}{:}& ~\text{No failure in the} ~ k \text{th measurement} \\
    H_1^{(k)}{:}& ~\text{A failure in the} ~ k \text{th measurement} \,.
\end{aligned}  
\label{eq:origin_hypo}
\end{equation}
The hypothesis testing for fault detection can be formalized by:\par
\noindent \textbf{Origin test}: $H_0^{(k)}$ is rejected if $|t_k|> \left(\mathbf{g}_k \mathbf{S}^{(k)}\mathbf{W}^{-1}\mathbf{S}^{(k)^T} \mathbf{g}_k^T + \sigma_k^2 \right)^{\frac{1}{2}} Q^{-1}(\frac{\alpha}{2})$ at significant level of $\alpha$, where $Q^{-1}(\cdot)$ is the quantile function of a standard normal variable. The probability of type I error (false alarm) of the origin test is $\alpha$. \par
In practice, the above test will be conducted for each subsolution to detect the potential failure in measurements, which evolves into a multiple-testing problem. In such a case, the type I error is actually enlarged. Thus, the following hypotheses are formalized instead, which are known as the Bonferroni correction \citep{bonferroni1936teoria}:
\begin{equation}
\begin{aligned}
    H_0{:}& ~\text{No failure in the} ~ n ~ \text{measurements} \\
    H_1{:}& ~\text{At least one failure in the} ~ n ~ \text{measurements} \,.
\end{aligned}    
\label{eq:correct_hypo}
\end{equation}
The hypothesis testing using the corrected hypotheses is formalized by:\par
\noindent \textbf{Corrected test}: $H_0$ is rejected if any $|t_k|> \left(\mathbf{g}_k \mathbf{S}^{(k)}\mathbf{W}^{-1}\mathbf{S}^{(k)^T} \mathbf{g}_k^T + \sigma_k^2 \right)^{\frac{1}{2}} Q^{-1}(\frac{\tau}{2n})$ at significant level of $\alpha^*$, where $\tau$ is the upper limit of $\alpha^*$. The probability of type I error (false alarm) of the corrected test is $\alpha^*$. \par
In implementing the corrected test, $\tau$ will be specified (e.g., 0.05) according to the nature of the application. Then the type I error $\alpha$ of the individual test would be $\frac{\tau}{n}$ (as shown in Appendix \ref{app:Bonferroni}), which could be very small when $n$ takes a large value. Therefore, the individual test and the corrected test both could be conservative. However, in satellite navigation applications, it is rare to have a large $n$, which ensures the feasibility of the corrected test.\par

\section{Relationships with Multiple Hypothesis Solution Separation}\label{sec:relation}
The separation between the full solution and the $k$th subsolution is given by \citep{brenner1996integrated, blanch2010raim,blanch2015baseline}
\begin{equation}
    d_{k,q} = (\hat{\mathbf{x}}- \hat{\mathbf{x}}^{(k)})_q, q=1,2,3 \,,
\label{eq:SS}
\end{equation}
where the subscript $q$ represents the $q$th component of the solution. By substituting Eqs. \eqref{eq:subsolution}, \eqref{eq:full solution}, and \eqref{eq:general_linear} into Eq. \eqref{eq:SS}, $d_{k,q}$ can be written as the linear combination of measurement errors:
\begin{equation}
    d_{k,q} = \sum\limits_{j=1}^{n} \tilde{s}_{q,j}\varepsilon_j \,,
    \label{eq:ss_linear}
\end{equation}
where $\tilde{s}_{q,j}$ is the element in the $q$th row and $j$th column of $\mathbf{S}-\mathbf{S}^{(k)}$.   \\

Define the separation vector $\mathbf{d}_k$ as follows:
\begin{equation}
    \mathbf{d}_k = [d_{k,1},d_{k,2},d_{k,3}]^T .
\end{equation}
Appendix \ref{app:relation} identifies the following relationship between $t_k$ and $\mathbf{d}_k$:
\begin{equation}
    \mathbf{d}_k = (\mathbf{G}^T \mathbf{W}\mathbf{G})^{-1} \mathbf{g}_k^T W_{k,k} t_k
    \label{eq:ss_jk_relation}
\end{equation}
Intuitively, this relationship reveals that the difference in solutions, induced by excluding the $k$th measurement, is proportional to the perturbation of the $k$th measurement and oriented along the direction $(\mathbf{G}^T \mathbf{W}\mathbf{G})^{-1} \mathbf{g}_k^T W_{k,k}$ in the solution space. Indeed, $(\mathbf{G}^T \mathbf{W}\mathbf{G})^{-1} \mathbf{g}_k^T W_{k,k}$ is the derivative of the solution $\hat{\mathbf{x}}$ with respect to the $k$-th measurement
\begin{equation}
    \frac{\diff \hat{\mathbf{x}}}{\diff y_k} = (\mathbf{G}^T \mathbf{W}\mathbf{G})^{-1} \mathbf{g}_k^T W_{k,k} \,,
\end{equation}
which quantifies how each element in $\hat{\mathbf{x}}$ changes in response to a small perturbation in $y_k$. Since the vector $(\mathbf{G}^T \mathbf{W}\mathbf{G})^{-1} \mathbf{g}_k^T W_{k,k}$ is uniquely defined for each $k$, there is a one-to-one correspondence between $\mathbf{d}_k$ and $t_k$, which demonstrates the equivalence of solution separation and jackknife residual. Given that solution separation represents the optimal detection statistic under single-fault conditions \citep{blanch2017theoretical}, it follows that the jackknife residual also possesses this optimality property. \par

In addition, the relationship in Eq. \eqref{eq:ss_jk_relation} indicates mapping from a scalar to a high-dimensional vector, revealing the difference in computation load between the SS and jackknife detectors during hypothesis testing. Specifically, the relationship in Eq. \eqref{eq:ss_jk_relation} shows that the statistical significance encapsulated in $t_i$ is allocated to each element of $\mathbf{d}_k$ through the vector $(\mathbf{G}^T \mathbf{W}\mathbf{G})^{-1} \mathbf{g}_k^T W_{k,k}$. This necessitates conducting the hypothesis test for each component of $\mathbf{d}_k$, requiring the determination of the nominal distribution for every element of $\mathbf{d}_k$. 
When the measurement error is modelled as Gaussian, this process remains tractable due to closed-form distributions; but under non-Gaussian measurement error modeling, this process needs to compute convolutions of non-Gaussian distributions for each element of $\mathbf{d}_k$, which is computationally expensive. However, the jackknife detector only needs to determine the distribution of a scalar $t_k$ by taking convolutions of non-Gaussian distributions. In a single-constellation system, the computational load of the SS detector is three times that of the jackknife detector.

\section{BOUNDING NON-GAUSSIAN NOMINAL ERRORS}\label{sec:ob_models}
To characterize and simplify the error profile of nominal measurements, the technique of overbound is widely adopted in the navigation communities \citep{braff2005method,shively_comparison_2001,larson_gaussianpareto_2019}. Overbound represents the worst possible error distribution in the absence of a hardware fault \citep{decleene_defining_2000,rife_core_2004}. Researchers have proposed various Gaussian, semi-Gaussian, and non-Gaussian overbounding methods. A review of these methods can refer to \citep{rife2012overbounding,yan2024principal}. This section will introduce two practical overbounding methods, including the Gaussian overbound \citep{decleene_defining_2000} and the Principal Gaussian overbound \citep{yan2024principal}. The latter one is a non-Gaussian overbounding method. Both methods will be used for bounding non-Gaussian nominal errors in Section \ref{sec:world simulation}.

\subsection{Gaussian overbound} \label{subsec:tsgo}
Let the cumulative distribution function (CDF) of the random variable $v$ be $G_v$. The Gaussian overbound is determined by finding the minimum $\delta$ that satisfies
\begin{subequations}
\begin{align}
    {\int_{- \infty}^{x}{f_\mathcal{N}\left( {x;0,\delta} \right)}}dx {\geq}& G_v(x)~\forall x < 0 \\
      {\int_{- \infty}^{x}{f_\mathcal{N}\left( {x;0,\delta} \right)}}dx {\leq}& G_v(x)~\forall x \geq 0 \,,
    \label{eq_TSGO_right}
\end{align}
\end{subequations}
where $f_\mathcal{N}(x;0,\sigma)$ is the PDF of a zero-mean Gaussian distribution with a standard deviation of $\sigma$.
\subsection{Principal Gaussian Overbound}\label{subsec:pgo}
The Principal Gaussian overbound \citep{yan2024principal} utilizes the zero-mean bimodal Gaussian mixture model (BGMM) to fit the error distribution based on the expectation–maximization (EM) algorithm \citep{dempster_maximum_1977} and divides the BGMM into the core and tail regions based on the analysis of BGMM membership weight. Within each region, one of the Gaussian components in the BGMM holds a dominant position, and a CDF overbound is constructed based on the dominant Gaussian component. The PDF of the Principal Gaussian overbound (PGO) is given by 
\begin{equation}
    f_{PGO}(x) = \begin{cases}
        \left( {1 + k} \right)\left( {1 - p_{1}} \right)f_N\left( {x;0,\sigma_{2}} \right) &|x| > x_{rp} \\
        p_{1}f_N\left( {x;0,\sigma_{1}} \right) + c &|x| \leq x_{rp} 
    \end{cases} \,,
\label{eq_PGO}
\end{equation}
where $f_N\left( {x;0,\sigma_{1}} \right)$ and $f_N\left( {x;0,\sigma_{2}} \right)$ are the PDF of the first and the second Gaussian component of the fitted BGMM, $\sigma_{1}$ and $\sigma_{2}$ are the corresponding standard deviations, and $p_1$ and $1-p_1$ are the mixing weight of the two Gaussian components, respectively; $k$, $c$, and $x_{rp}$ are parameters uniquely determined by the partition strategy based on the analysis of BGMM membership weight \citep{yan2024principal}. The variance of PGO can be calculated by conducting integration over $x$ as follows:
\begin{equation}
    \Var(x) = \int_{-\infty}^{\infty} x^2 f_{PGO}(x) dx \,.
\label{eq:var_pgo}
\end{equation}
A detailed description of PGO can refer to \cite{yan2024principal}. Soon, it will be shown in Section \ref{sec:world simulation} that PGO provides a sharper yet conservative overbound than the Gaussian overbound for heavy-tailed error distribution. Notably, it is proved that PGO can maintain the overbounding property through convolution \citep{yan2024principal}, which is the basis for deriving pseudorange-level requirements from the position domain integrity requirements \citep{decleene_defining_2000}. \par

\section{DETECTION PERFORMANCE WITH WORLDWIDE SIMULATIONS}\label{sec:world simulation}
This section shows the single-failure detection results of a set of users distributed over the world during one day. MAAST \citep{jan2001matlab}, a MATLAB toolset developed at Stanford University, is utilized to simulate the dual frequency pseudorange measurements, satellite positions, and user locations. Specifically, the 27-satellite Galileo constellation \citep{zandbergen2004galileo} is used to simulate satellite positions. The users are placed on a grid every 10 degrees longitude and latitude (which gives 648 locations). For each location, the geometries are simulated every 5 min (which gives 288 time steps). The dual pseudorange frequency measurements are simulated by adding randomly generated samples from a given error distribution to the true range. The measurement model is given by dual-frequency SPP, as shown in Appendix \ref{app:linearform}. For each time and user location, an artificial bias (\SI{10}{m}) is injected into one of the measurements randomly. The SS detector and the proposed jackknife detector are implemented to detect these faults separately. The type I error (false alarm rate) for both detectors is set as 0.05. The detection rate at each user location is defined as
\begin{equation}
    P_{dec} = \frac{\text{Detected epochs in one day}}{\text{Valid epochs in one day}} \,,
\label{eq:detect rate}
\end{equation}
where the denominator could be less than 288 since the number of satellites in view may not satisfy the minimum requirements for fault detection (for SPP, the minimum number is 5).\par

\subsection{Nominal error simulation and bounding}\label{sec:NIG_error}
The NIG distribution is found to be useful in approximating the heavy tails of the ground-station error distribution for LAAS \citep{braff2005method, rife_core_2004}, and its PDF is given by
\begin{equation}
    f_{NIG}(x) = \frac{\delta_0^2 \exp{(\delta_0^2)}}{\pi \sqrt{x^2+\delta_0^2}}K_1\Big(\delta_0\sqrt{x^2+\delta_0^2}\Big) \,,
\end{equation}
where $\delta_0$ is the shape parameter that determines the weight of the NIG tail, and $K_1$ is a modified Bessel function of the second kind, degree one. Similar to the setting in \cite{braff2005method}, $\delta_0=0.65$ is adopted to simulate the measurement error $\varepsilon_i$. Note that the error distribution for each measurement is set to be identical. The primary reason for this simplification is that, to the best of the authors’ knowledge, there is no developed method for establishing the geometry-related NIG error model. Since the simplification is applied for both detectors, its impact on the conclusion drawn from the subsequent experimental results is expected to be negligible.\par

\begin{figure}[!htb]
    \centering
  \subfloat[~~~~]{%
       \includegraphics[width=0.3\textwidth]{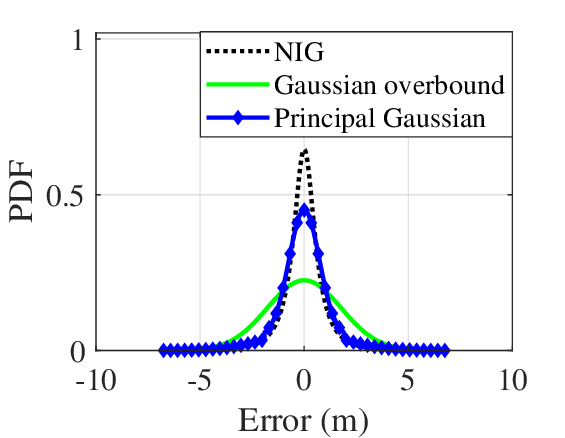}}
  \subfloat[~~~~]{%
        \includegraphics[width=0.3\textwidth]{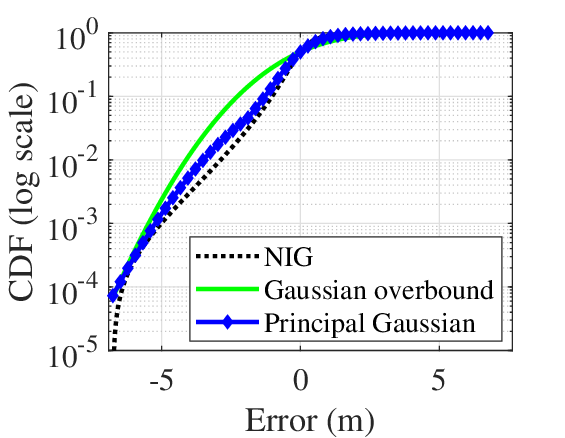}}
  \caption{The (a) PDF and (b) CDF (plotted in logarithmic scale) of the Gaussian overbound and Principal Gaussian overbound for the NIG distribution ($\delta_0=0.65$).}
  \label{fig:PGO_TSGO_NIG}
\end{figure}

The Gaussian overbound \citep{decleene_defining_2000} and PGO \citep{yan2024principal} illustrated in Section \ref{sec:ob_models} are employed to bound the NIG distributed measurement error. The weighting matrix for the jackknife detector is constructed as a diagonal matrix, with each diagonal element taking the inverse of the variance of the Gaussian overbound or PGO.  Fig. \ref{fig:PGO_TSGO_NIG} plots the Gaussian overbound and the PGO of the NIG distribution ($\delta_0=0.65$). As can be seen, PGO provides a sharper bound for the NIG distribution at both the tail and core regions than the Gaussian overbound. 

\subsection{Detection Performance Analysis}\label{sec:detect_NIG}
\begin{figure}[!htb]
    \centering
  \subfloat[~~~~]{%
       \includegraphics[width=0.4\textwidth]{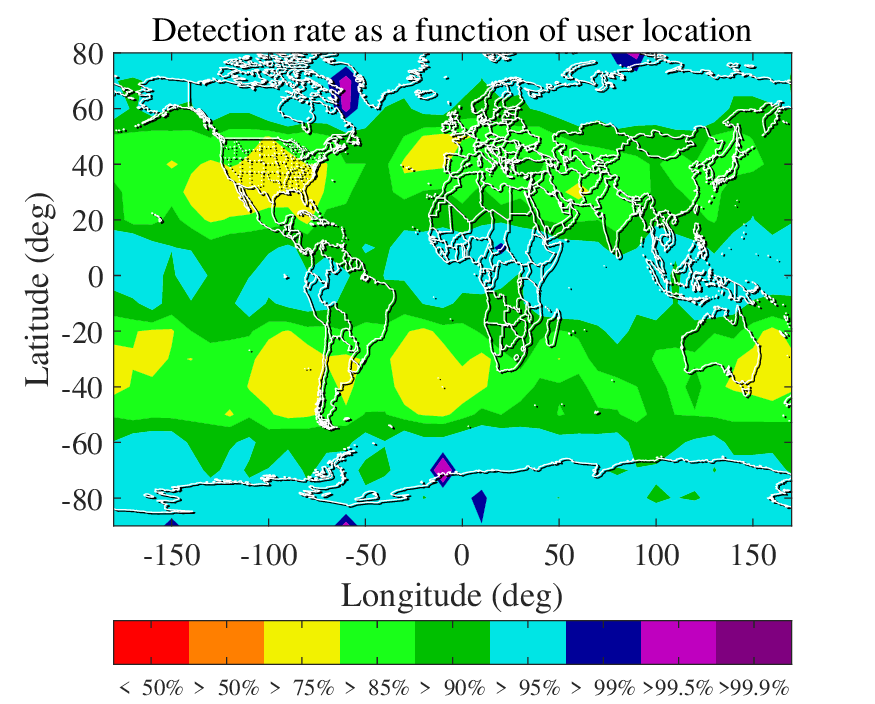}}
  \subfloat[~~~~]{%
        \includegraphics[width=0.4\textwidth]{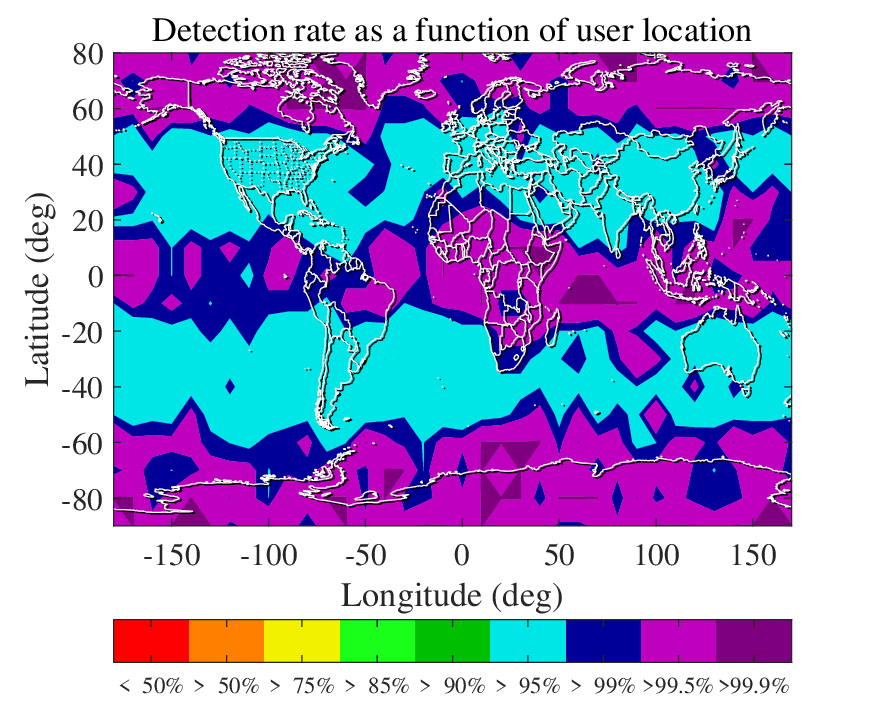}}
    \\
  \subfloat[~~~~]{%
       \includegraphics[width=0.4\textwidth]{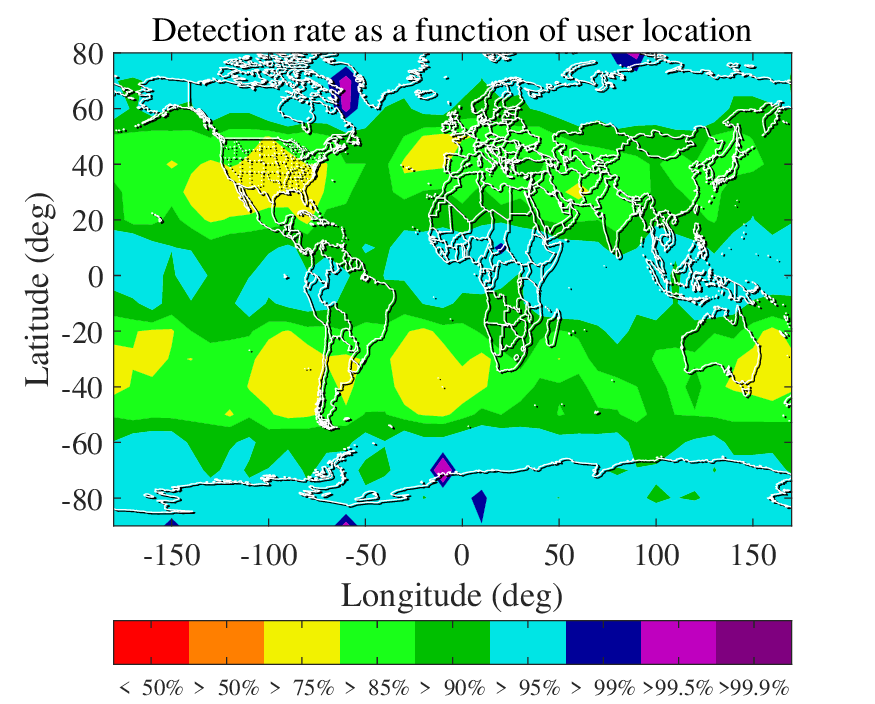}}
  \subfloat[~~~~]{%
        \includegraphics[width=0.4\textwidth]{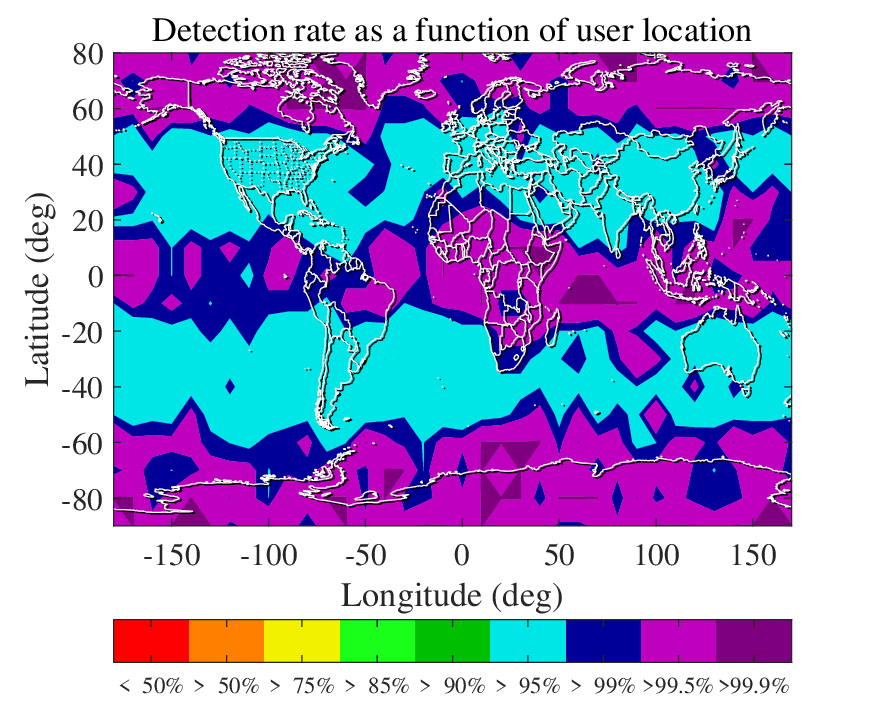}}
  \caption{Detection performance of the jackknife detector with artificially injected bias (\SI{10}{\meter}) under NIG distributed measurement errors when using (a) the Gaussian overbound; b) PGO for nominal error bounding; Detection performance of the SS detector in the same setting when using (c) the Gaussian overbound; d) PGO for nominal error bounding.}
  \label{fig:world_NIG}
\end{figure}

The top panel of Fig. \ref{fig:world_NIG} shows the contour plot of the detection rate for the jackknife detector using different overbounding methods, revealing a noticeable disparity. The jackknife detector using the PGO exhibits a substantial enhancement in detection rate when compared to that using the Gaussian overbound. In most user locations, the jackknife detector using the PGO achieves a detection rate of over \SI{95}{\percent}. Moreover, in a considerable number of user locations, the detection rate even surpasses \SI{99.5}{\percent}. However, the jackknife detector using the Gaussian overbound exhibits a detection rate of \SI{85}{\percent} at nearly half of the locations. The primary reason is that the sharper overbound provided by PGO can better characterize the measurement error distribution than the Gaussian overbound, as shown in Fig. \ref{fig:PGO_TSGO_NIG}. The PGO provides an accurate probabilistic model for hypothesis testing, which guarantees the performance of hypothesis testing. These results demonstrate the benefits of introducing non-Gaussian overbounds into the jackknife detector. The bottom panel of Fig. \ref{fig:world_NIG} shows the detection results of the SS detector. Evidently, the SS detector demonstrates identical performance to the jackknife detector using corresponding overbounding methods. These results support the findings in Section \ref{sec:relation}. \par

\section{DETECTION PERFORMANCE THROUGH MONTE CARLO SIMULATION}\label{sec:MC}
Section \ref{sec:world simulation} uses a worldwide simulation to demonstrate the equivalent detention performance of the jackknife and SS detectors regardless of geometry. However, at each location-time (i.e., a fixed geometry), the detection performance is only evaluated on one set of measurements, which means the detection results depend on the current set of instances of the measurement noises (random variables). To fully characterize the detection performance of both detectors, a Monte Carlo simulation is necessary to reveal the impact of the stochastic nature of measurement noises on the detection results, which is the main focus of this section.\par
\begin{figure}[!htb]
    \centering
  \subfloat[~~~~]{%
       \includegraphics[width=0.4\textwidth]{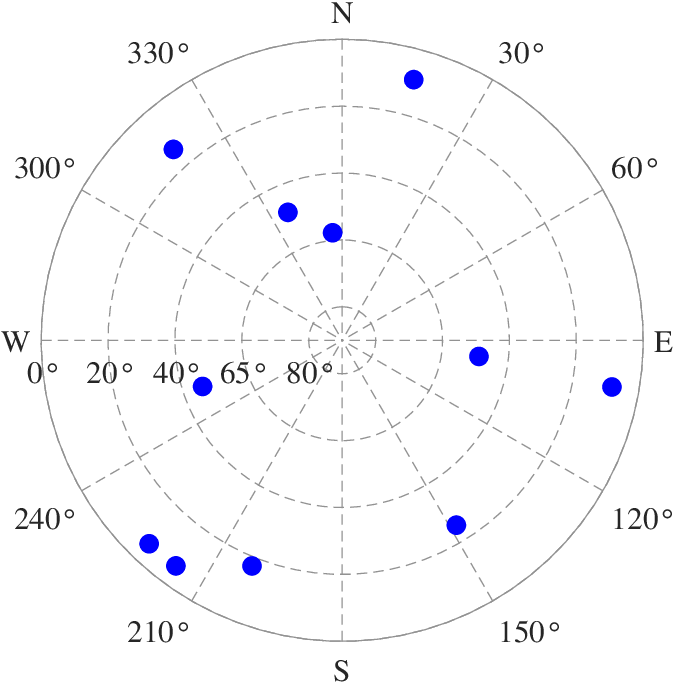}}
  \subfloat[~~~~]{%
        \includegraphics[width=0.5\textwidth]{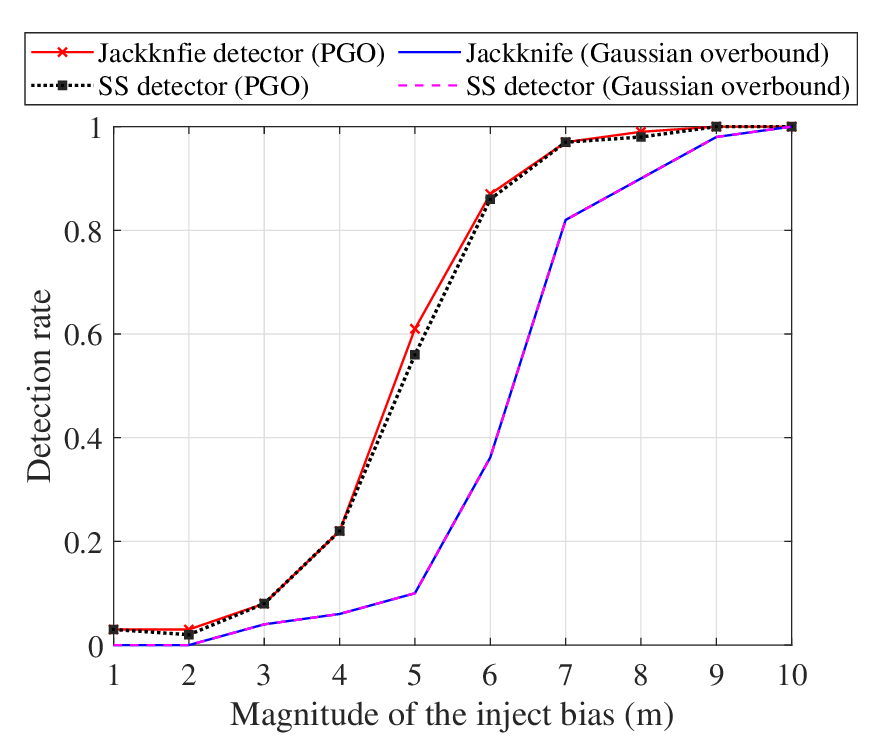}}
  \caption{The sky plot of the satellite geometry. (b) The detection rate of the jackknife and SS detectors with different magnitudes of the injected bias.}
  \label{fig:MC_decection}
\end{figure}

We randomly select a location-time pair from the worldwide simulated dataset in Section \ref{sec:world simulation}, and the satellite geometry is shown in Fig. \ref{fig:MC_decection}a. Similar to the settings in Section \ref{sec:NIG_error}, we use the NIG distribution ($\sigma_0=0.65$) to simulate the measurement error $\varepsilon_i$ , and set the error distribution for each measurement as identical. Both the Gaussian overbound and the PGO are adopted to bound the NIG distributed measurement error. Ten experiments are performed to examine the detection performance of the jackknife and SS detectors with different overbounding methods, each involving the injection of bias with a specific magnitude (ranging from 1m to 10m with a step size of 1m). In each experiment, we conduct 1,000 Monte Carlo simulations, where an artificial bias is injected into one of the measurements in each run. \par

Fig. \ref{fig:MC_decection}b shows the detection rate of each method. In all cases, the detectors using the PGO significantly outperform those using the Gaussian overbound. The jackknife and SS detectors yield the same results when both of them use the Gaussian overbound. Interestingly, when the magnitude of the inject bias is 2m, 5m, 6m, 8m, the jackknife detector using the PGO exhibits a slightly higher detection rate than that of the SS detector using the PGO. Theoretically, based on the findings of Section \ref{sec:relation}, the SS detector should demonstrate the same performance as the jackknife detector if both of them use the PGO for nominal error bounding. These occasional discrepancies between the performance of the jackknife and SS detectors can be attributed to numerical instability inherent in non-Gaussian convolution. This instability is further amplified by the sensitivity of hypothesis testing to tail probability estimates. While jackknife and SS detectors are theoretically equivalent, their numerical implementations accumulate errors differently (Eqs. \eqref{eq:ti_3} and \eqref{eq:ss_linear}), leading to rare disagreements near the decision boundary.

\section{Application to Real-world Satellite Clock Anomaly Detection}\label{sec:realworld}
The analysis conducted in Section \ref{sec:world simulation} primarily relies on simulated data. To showcase the practical applicability of the jackknife detector, this section focuses on evaluating its performance in real-world satellite clock anomaly detection. 

\subsection{Data Prepossessing and Overbounding}
GPS PRN-1 experienced a clock anomaly on January 28th, 2023, where the anomaly began at GPS time 15:02:30 and was set unhealthy at 16:05:00 \citep{lai2023prototyping}. The clock anomaly began to be corrected by manual control at 18:00, and the correction was completed at 20:00:00. The anomaly happened when PRN-1 was over in the middle of South Pacific, so the data collected from the nearby station, CHTI, located in Chatham Island, New Zealand, are used for analysis. Specifically, the observation data from January 1st, 2023, to January 31st, 2023 are collected from the Continuously Operating Reference Stations (CORS) website run by the National Geodetic Survey (NGS)\citep{NGS}. The position and receiver clock bias of the CHTI station at each epoch is solved using precise point positioning (PPP), with the fixed rate larger than \SI{98}{\percent}. The satellite position is calculated based on the broadcast ephemeris from NASA's Archive of Space Geodesy Data website \citep{NASA} by utilizing RTKLIB \citep{takasu2009development}. The pseudorange measurement is corrected by clock bias correction, ionospheric corrections, and tropospheric corrections based on RTKLIB \citep{takasu2009development} with the broadcast ephemeris. \par
\begin{figure}[!htb]
\centering
\includegraphics[width=\textwidth]{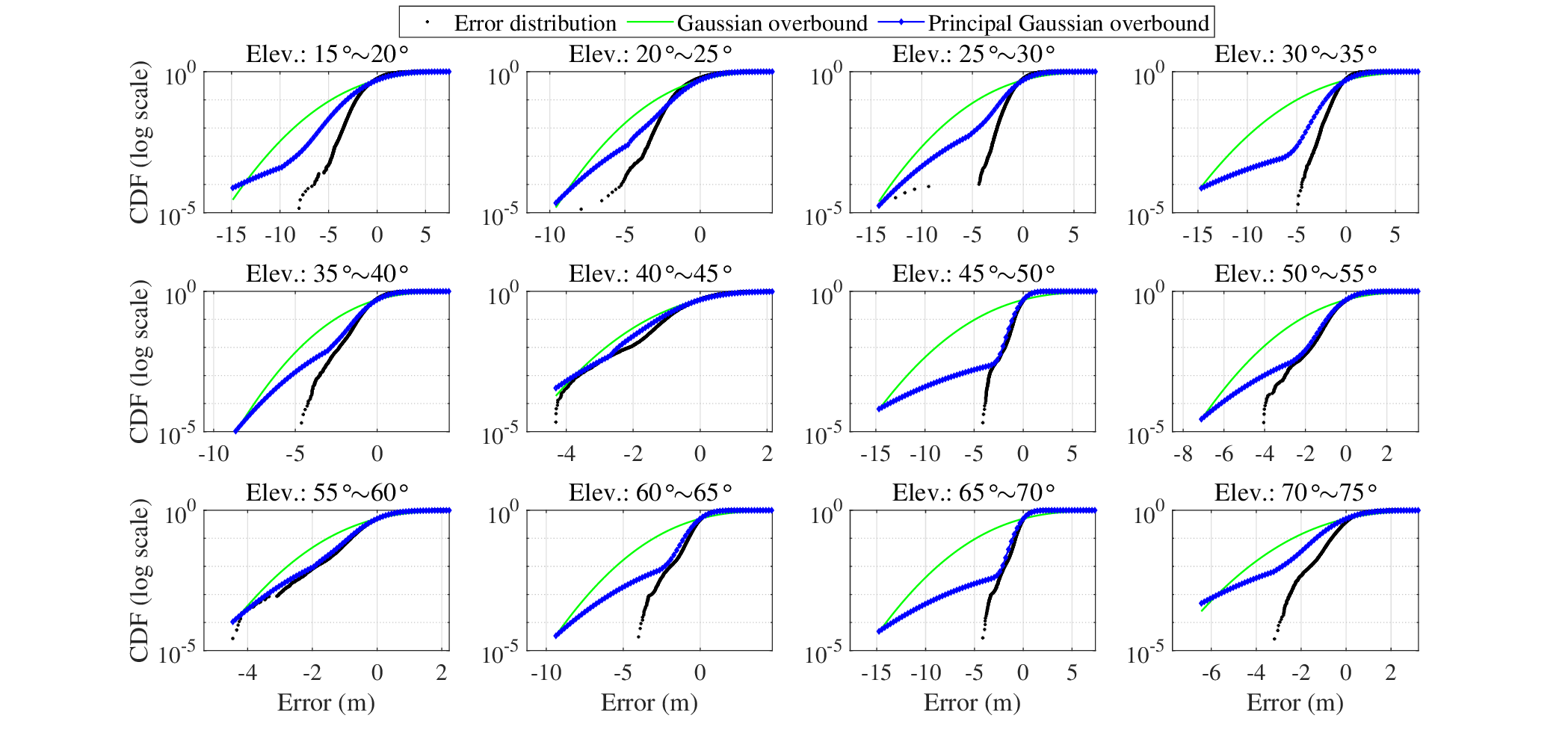}
\caption{The CDF of the Gaussian overbound and Principal Gaussian overbound for corrected pseudorange errors in each elevation angle bin. The CDF is plotted in logarithm scale.}
\label{fig:CHTI_CDF_compare}
\end{figure}

The pseudorange measurement error is calculated by subtracting the true range and receiver clock bias equivalent range from the corrected pseudorange measurement (only L1 measurements are studied, and faulty measurements are pre-excluded). Since GPS data are strongly influenced by the elevation angle, the pseudorange measurement error are organized into bins based on elevation angles every \SI{5}{\degree} from \SI{15}{\degree} to \SI{75}{\degree}, encompassing the highest observed elevation angle in the one-month dataset. Within each bin, the Gaussian overbound and the PGO for the pseudorange measurement error are computed and plotted in Fig. \ref{fig:CHTI_CDF_compare}. As can be seen, the PGO yields a sharper bound than the Gaussian overbound in all cases, where the measurement error shows significant heavy tails. Similar results are found in the complementary cumulative distribution function (CCDF) plot of the Gaussian overbound and the PGO for DGNSS errors, which are presented in Appendix \ref{app:CHTI_bounding}.\par

\subsection{Detection Performance Analysis}
\begin{figure}[!htb]
    \centering
  \subfloat[~~~~]{%
       \includegraphics[width=0.49\textwidth]{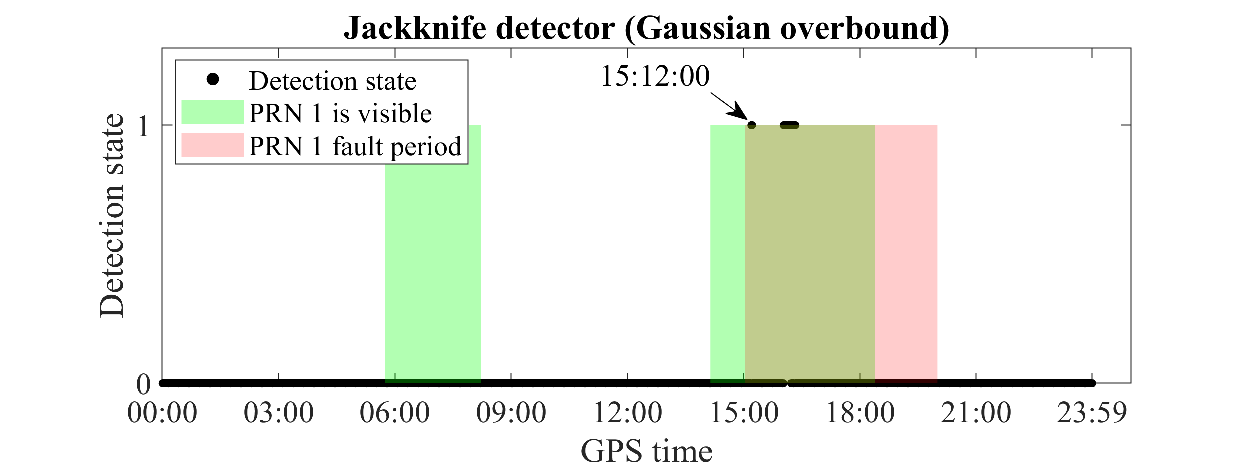}}
  \subfloat[~~~~]{%
        \includegraphics[width=0.49\textwidth]{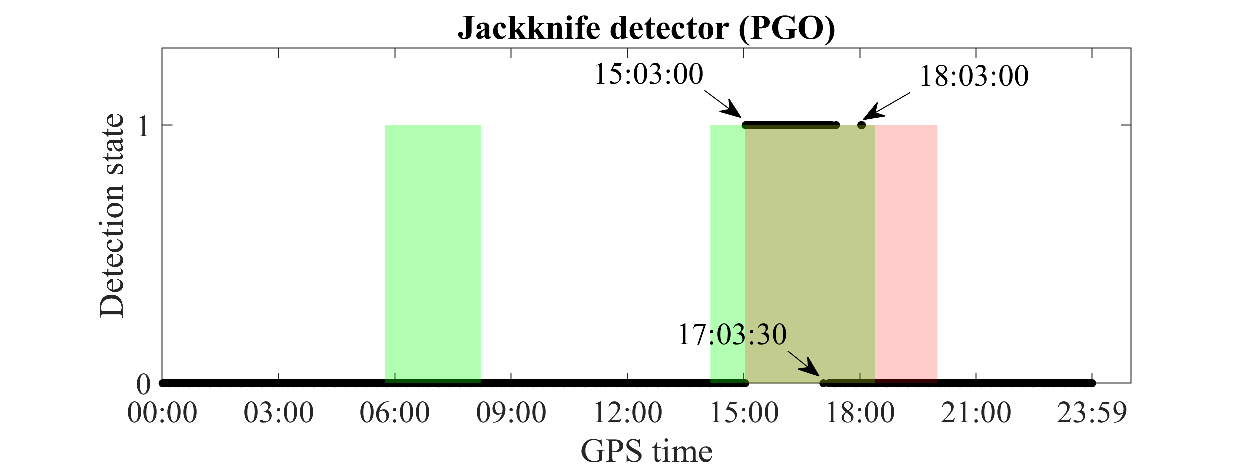}}
  \\
  \subfloat[~~~~]{%
       \includegraphics[width=0.49\textwidth]{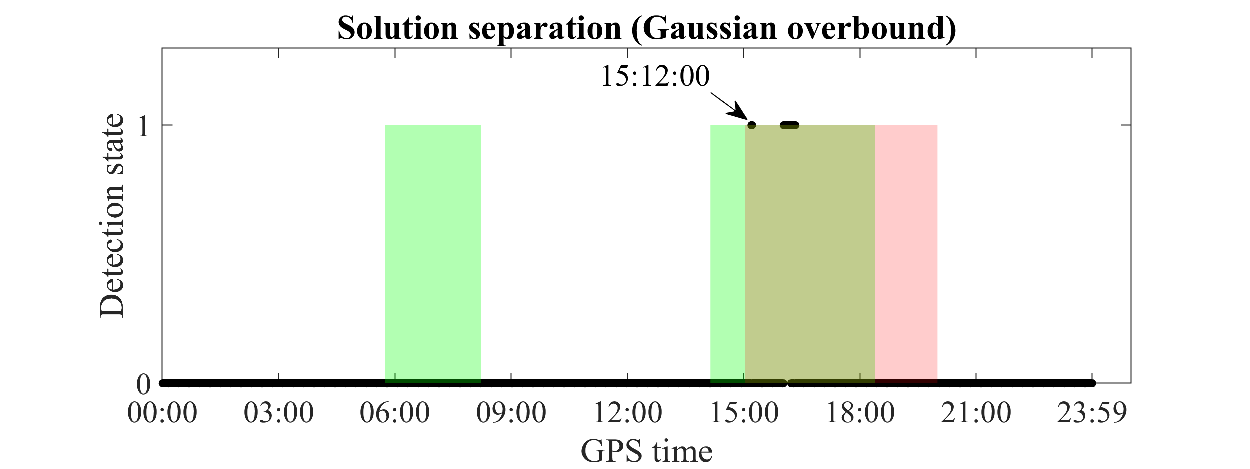}}
  \subfloat[~~~~]{%
        \includegraphics[width=0.49\textwidth]{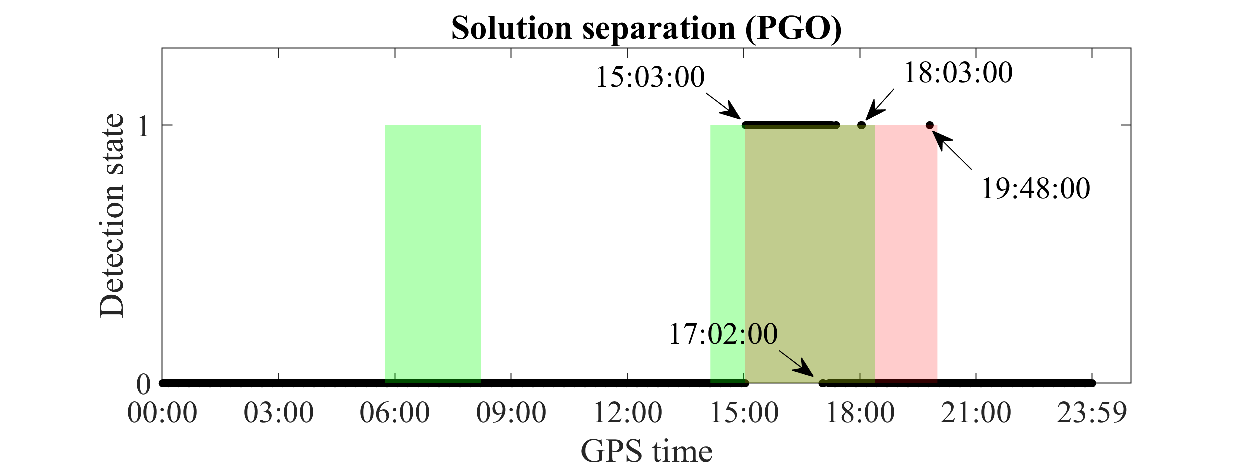}}
  \caption{Detection states of the jackknife detector in the clock anomaly detection task when using (a) the Gaussian overbound; b) PGO for nominal error bounding; Detection states of the SS detector in the same task when using (c) the Gaussian overbound; d) PGO for nominal error bounding; The detection state of "1" refers to "a fault is claimed" while "0" refers to "no failure is claimed".}
  \label{fig:Jan28}
\end{figure}

The jackknife and SS detectors are employed to detect the clock anomalies. For each detector, both the Gaussian overbound and the PGO are adopted to bound the nominal errors. Fig. \ref{fig:Jan28} shows the detection states of these two detectors using different overbounding methods at the CHTI station on January 28th. The type I error (false alarm rate) for all detectors is set as 0.05. PRN-1 is visible for CHTI in two time periods, i.e., 05:44:30$\sim$08:13:30 and 14:08:30$\sim$18:23:00, as marked by the green shaded area in Fig. \ref{fig:Jan28}. In addition, the fault period 15:02:30$\sim$20:00:00 is marked by the red shaded area. 

The left panel of Fig. \ref{fig:Jan28} shows the detection rate of the jackknife and SS detectors using the Gaussian overbound, where both detectors yield the same performance with the first claim of faults at 15:12:00. However, during the PRN 1 fault period, both detectors show extremely high missed detection rates. This situation is largely alleviated by introducing the non-Gaussian overbound, PGO, into the detectors, as shown in the right panel of Fig. \ref{fig:Jan28}. Both detectors can detect anomalies at most times of the fault period. Remarkably, these detectors claim the fault at 15:03:00, merely 30 seconds after the anomaly happened, which is 8 minutes earlier than the detectors using the Gaussian overbound. Since the sampling rate of the observation data retrieved from the CORS website is 30 seconds, the minimum delayed time of detection is limited to 30 seconds. With higher frequency observation data, detectors using non-Gaussian overbounds could yield a smaller delayed time of detection. \par

It is worth noting that a considerably long period of miss detection also exists in detectors using the PGO for nominal error bounding. Specifically, the miss-detection period began at 17:03:30 and lasted until 18:02:30 for the jackknife detector, during which the jackknife detector occasionally claimed faults.  The primary reason can be attributed to the inaccuracy of the nominal error model. Since only one-month data are used to model the pseudorange measurement error and compute the overbounds, the nominal error model could be conservative and can even shield small faults. Therefore, the detection performance of the jackknife detector could be degraded. \par

Fig. \ref{fig:Jan28}d shows that the miss-detection period of the SS detector began at 17:02:00, which is \SI{90}{\second} earlier than the jackknife detector. Moreover, the SS detector claims a fault at 19:48:00, during which PRN 1 is invisible. Therefore, the SS detector actually raised a false alarm at 19:48:00. Similar to the findings in the Monte Carlo simulations in Section \ref{sec:MC}, these occasional discrepancies between the performance of the jackknife and SS detectors can be attributed to numerical instability inherent in non-Gaussian convolution. 

\begin{figure}[!htb]
\centering
\includegraphics[width=0.8\textwidth]{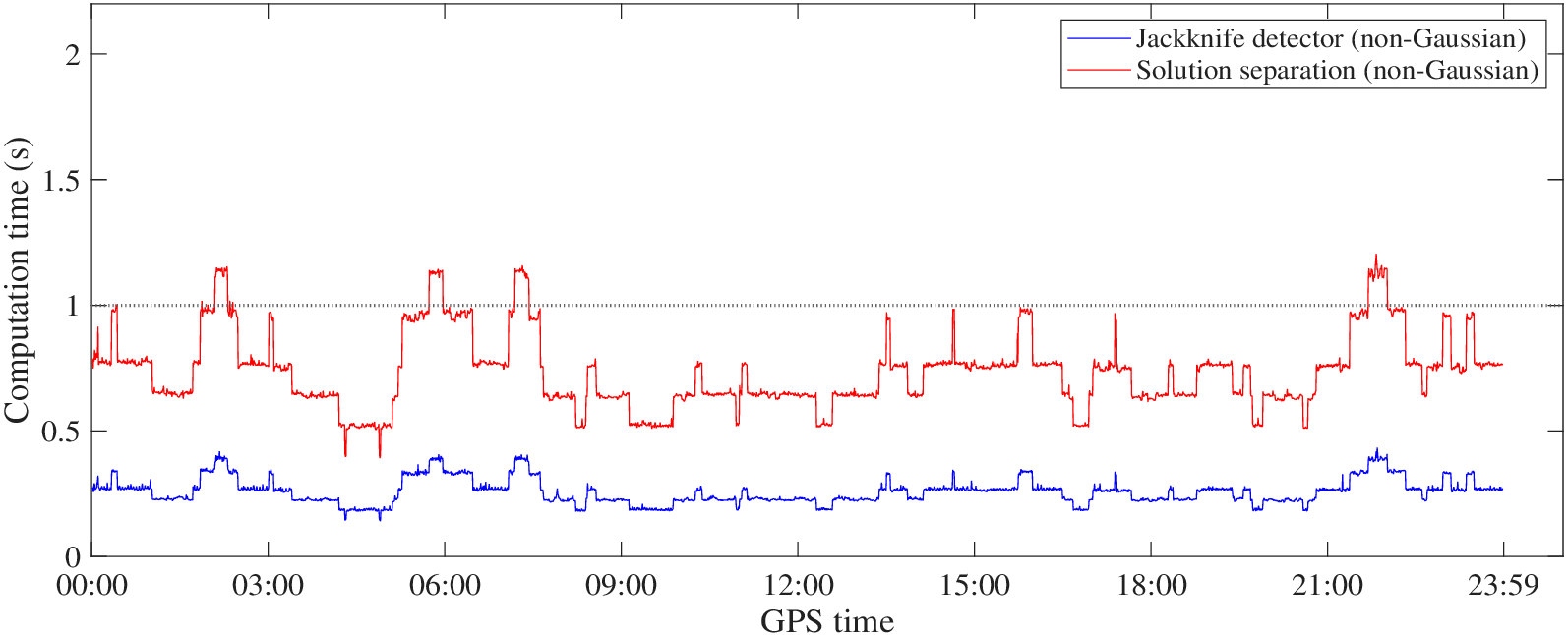}
\caption{The computation time of the jackknife and SS detectors when using the PGO for nominal error bounding in the clock anomaly detection task.}
\label{fig:time}
\end{figure}

Fig. \ref{fig:time} shows the computation time of the jackknife and SS detectors when using the PGO for nominal error bounding, where all computations are conducted on a laptop equipped with an Intel Core i7-12700H CPU running at 2.30GHz. As can be seen, the computation time for the jackknife detector remained below \SI{0.5}{\second} across all epochs. In contrast, as anticipated, the computation time of the SS detector was approximately threefold that of the jackknife detector. In a several epochs, the SS detector's computation time surpassed \SI{1}{\second}. For real-time applications utilizing commercial GNSS receivers, which typically have a sampling rate of 1 Hz, the jackknife detector offers an advantage owing to its computational efficiency.

\section{CONCLUSIONS AND FUTURE WORK}
This paper proposes a statistically rigorous and computationally efficient fault detection method for linearized pseudorange-based positioning systems under non-Gaussian nominal errors. Rather than introducing a fundamentally new detection algorithm, our primary contribution lies in providing a more efficient implementation that maintains statistical rigor while significantly reducing computational burden in non-Gaussian environments. Specifically, this paper constructs the jackknife residual by computing the inconsistency between the observed measurement and the predicted measurement based on subset solutions. Using the jackknife residual as the test statistic, a jackknife detector is developed by formalizing a multiple-testing problem with the Bonferroni correction to detect faults. It is proven that the constructed test statistic is the linear combination of measurement errors without making assumptions about the distribution of errors, which provides an accurate probabilistic model for hypothesis testing and establishes theoretical foundations for fault detection. We further establish the relationship between the jackknife and SS detectors, where the SS detector is proven to be the projection of the jackknife detector along the direction defined by the derivative of the solution with respect to the suspected measurement. This relationship reveals that while both methods are mathematically equivalent for single-fault detection, the jackknife detector requires only scalar statistical analysis compared to the vector-based computations needed for SS detection, resulting in significantly lower computational load.\par
In a worldwide simulation with $648\times288$ location-time geometries, we demonstrate the equivalent performance between the jackknife detector and the SS detector. We also showcase the benefits of introducing non-Gaussian overbounds into these detectors. 
Additionally, Monte Carlo analysis with 1,000 simulations provides rigorous statistical validation of the theoretical equivalence between jackknife and SS detectors. While both methods demonstrate comparable detection performance, occasional minor discrepancies (the jackknife detector occasionally shows marginally better detection performance than the SS detector when nominal error models are non-Gaussian) highlight the sensitivity of hypothesis testing to numerical implementation differences in non-Gaussian distribution handling. Moreover, this paper investigates a real-world application of detecting clock anomalies of PRN-1 on January 28th, 2023. The jackknife detector using the non-Gaussian overbound demonstrates an extremely short detection delay of \SI{30}{\second}, which is 8 minutes less than the jackknife detector with the Gaussian overbound. As expected, the SS detector with the non-Gaussian overbound yields the same outcome. However, the jackknife detector achieves a consistent threefold improvement in computational efficiency compared to the SS detector, making it particularly valuable for real-time safety-critical applications where computational resources are limited. \par
The proposed method provides an efficient implementation for detecting faults in linearized pseudorange-based positioning systems under non-Gaussian nominal errors while maintaining the statistical optimality properties inherent in solution separation methods. However, the proposed method is mainly designed for single-fault cases. For multiple-fault detection under non-Gaussian nominal errors, a possible solution could be examining the most influential measurements with combinatorial jackknife testing, which will be the main focus of our future work. Moreover, the proposed method assumes that the nominal measurement error has a zero mean. A systematic way to consider the bias effect on the fault detection results is needed and remains for further exploration. 

\printbibliography[title=References]

\appendix
\renewcommand{\thesection}{Appendix \Alph{section}}
\newpage

\section{LINEARIZATION FOR SINGLE POINT POSITIONING (SPP)}\label{app:linearform}
The right-hand side (RHS) of the pseudorange measurement model in Eq. \eqref{eq:meas_model} can be linearized by taking the first-order Taylor expansion at $x_{0}=\left[u_{x,0},u_{y,0},u_{z,0}, u_{t,0}\right]^T$ as follows:
\begin{equation}
\begin{aligned}
    \varrho_i {=}& \varrho_{i,0} -a_{i,1}\left(u_x-u_{x,0}\right)-a_{i,2}\left(u_y-u_{y,0}\right)\\
    &{-}a_{i,3}\left(u_z-u_{z,0}\right) + u_t - u_{t,0} +\eta_i \,,
\label{eq:linear_meas_model}
\end{aligned}
\end{equation}
where 
\begin{subequations}
\begin{align}
    \varrho_{i,0}{=}&\sqrt{\left(p_{x}^{i}-u_{x,0}\right)^2+\left(p_{y}^{i}-u_{y,0}\right)^2+\left(p_{z}^{i}-u_{z,0}\right)^2}+u_{t,0} \\
    a_{i,1}{=}&\frac{p_{x}^{i}-u_{x,0}}{\sqrt{\left(p_{x}^{i}-u_{x,0}\right)^2+\left(p_{y}^{i}-u_{y,0}\right)^2+\left(p_{z}^{i}-u_{z,0}\right)^2}} \\
    a_{i,2}{=}&\frac{p_{y}^{i}-u_{y,0}}{\sqrt{\left(p_{x}^{i}-u_{x,0}\right)^2+\left(p_{y}^{i}-u_{y,0}\right)^2+\left(p_{z}^{i}-u_{z,0}\right)^2}} \\
    a_{i,3}{=}&\frac{p_{z}^{i}-u_{z,0}}{\sqrt{\left(p_{x}^{i}-u_{x,0}\right)^2+\left(p_{y}^{i}-u_{y,0}\right)^2+\left(p_{z}^{i}-u_{z,0}\right)^2}}\,.
\end{align}
\end{subequations}
The matrix form of the linearized pseudorange measurement model with $n$ measurements can be written as
\begin{equation}
    y=\mathbf{G}x+\varepsilon \,,
    \label{eq:SPP matrix form}
\end{equation}
where
\begin{equation}
\begin{aligned}
    &y=\begin{bmatrix}
        \varrho_{1,0}-\varrho_1 \\
        \vdots \\
        \varrho_{n,0}-\varrho_n
    \end{bmatrix}\,,
    \mathbf{G} = \begin{bmatrix}
     a_{1,1} & a_{1,2} & a_{1,3} & 1\\
     \vdots  & \vdots  & \vdots & \vdots \\
    a_{n,1} & a_{n,2} & a_{n,3}  & 1
    \end{bmatrix} \,, \\
     &\varepsilon=\begin{bmatrix}
        \eta_1 \\
        \vdots \\
        \eta_n
    \end{bmatrix} \,,
    x=\begin{bmatrix}
        u_x-u_{x,0} \\
        u_y-u_{y,0} \\
        u_z-u_{z,0} \\
        -(u_t - u_{t,0})
    \end{bmatrix}\,.
\label{eq:SPP matrix meaning}
\end{aligned}
\end{equation}
For single-frequency receivers (such as L1 or L5), the measurement error $\eta_i$ mainly includes the satellite clock and ephemeris error $\epsilon_{eph\&clk,i}$, ionospheric delay $\epsilon_{iono,i}$, tropospheric delay $\epsilon_{trop,i}$ and code noise and multipath error $\epsilon_{cnmp,i}$:
\begin{equation}
    \eta_i = \epsilon_{eph\&clk,i} + \epsilon_{iono,i} + \epsilon_{trop,i} + \epsilon_{cnmp,i}\,.
\end{equation}
For double frequency receivers (e.g., L1-L5), the ionospheric delay can be eliminated by differencing pseudorange measurements on both frequencies as follows \citep{misra_global_2006}:
\begin{subequations}
\begin{align}
    \varrho_{iono\text{-}free}=\frac{\gamma \varrho_{L1}-\varrho_{L5}}{\gamma-1} \\\
    \gamma = \left(\frac{f_{L1}}{f_{L5}}\right)^2 \,,
\end{align}
\label{eq:iono-free pseudorange}
\end{subequations}
where $\varrho_{iono\text{-}free}$ represents the ionospheric-free pseudorange, and $f_{L1}$ and $f_{L5}$ represents the L1 and L5 frequencies, respectively. Then, the measurement error can be represented as follows \citep{blanch2023evaluation}:
\begin{equation}
    \eta_i = \epsilon_{eph\&clk,i}^{L1\text{-}L5} + \epsilon_{trop,i} + \frac{\gamma}{\gamma-1}\epsilon_{cnmp,i}^{L1} - \frac{1}{\gamma-1}\epsilon_{cnmp,i}^{L5} \,,
    \label{eq:dual freq error}
\end{equation}
where $\epsilon_{eph\&clk,i}^{L1\text{-}L5}$ is the satellite clock and ephemeris error for the L1-L5 ionospheric free combination. With the definition in Eqs. \eqref{eq:iono-free pseudorange} and \eqref{eq:dual freq error}, the linearized pseudorange measurement model for the dual frequency SPP takes a similar form as in Eqs. \eqref{eq:SPP matrix form} and \eqref{eq:SPP matrix meaning}.

\section{DISTRIBUTION OF JACKKNIFE RESIDUAL UNDER GAUSSIAN NOISES}\label{app:JK_res_proof}
The Gauss-Markov conditions concern the set of noises in the linear system $y=\mathbf{G}x+\bm{\varepsilon}$ as follows:
\begin{enumerate}
    \item Zero mean: $\E[\varepsilon_i] = 0 ~\forall i$;
    \item Homoscedastic: $\Var[\varepsilon_i]=\sigma^2<\infty~\forall i$;
    \item Uncorrelated: $\Cov[\varepsilon_i,\varepsilon_j]=0 ~\forall i\neq j$.
\end{enumerate} 
Under Gauss-Markov conditions, the ordinary least squares (OLS) estimator is the best linear unbiased estimator (BLUE). \par
A further generalization of the Gauss-Markov conditions to heteroscedastic and correlated errors has been developed \cite{aitken1936on}, and its application to the weighted least squares (WLS) estimator can be stated as follows:
\begin{displayquote}
``WLS is the BLUE if the weight matrix is equal to the inverse of the variance-covariance matrix of the noises."
\end{displayquote}
Based on the generalized Gauss-Markov conditions, the subsolution in Eq. (\ref{eq:subsolution}a) has the following properties:
\begin{subequations}
\begin{align}
    \E[\hat{\mathbf{x}}^{(k)}]{=}&0 \\
    \Var[\hat{\mathbf{x}}^{(k)}]{=}&\mathbf{S}^{(k)}\mathbf{W}^{-1}\mathbf{S}^{(k)^T} \,.
\end{align}
\end{subequations}
By substituting Eq. \eqref{eq:prediction} into Eq. \eqref{eq:JK_residual}, the jackknife residual can be written by
\begin{equation}
\begin{aligned}
    t_k &= y_k - \mathbf{g}_k \hat{\mathbf{x}}^{(k)} \\
    & = \mathbf{g}_k \mathbf{x}^{(k)} + \varepsilon_k  - \mathbf{g}_k \hat{\mathbf{x}}^{(k)} \\
    & = \mathbf{g}_k (\mathbf{x}^{(k)}-\hat{\mathbf{x}}^{(k)}) + \varepsilon_k \,.
\end{aligned}
\label{eq:ti_1}
\end{equation}
The expectation and variance of the jackknife residual in \eqref{eq:ti_1} are given by
\begin{subequations}
\begin{align}
    \E[t_k] {=}& 0 \\
    \Var[t_k] {=}& \mathbf{g}_k \Var\left[  \mathbf{x}^{(k)} - \hat{\mathbf{x}}^{(k)}\right] \mathbf{g}_k^T + \sigma_k^2 \notag \\
            {=}& \mathbf{g}_k \mathbf{S}^{(k)}\mathbf{W}^{-1}\mathbf{S}^{(k)^T} \mathbf{g}_k^T + \sigma_k^2 \,.
\end{align}
\end{subequations}
As shown in Eq. \eqref{eq:ti_2}, the Jackknife residual can be rewritten as
\begin{equation}
    t_k= \tilde{\mathbf{p}}_k \bm{\varepsilon} \,,
\end{equation}
which is a linear combination of measurement errors. If $\varepsilon_i$ has a zero-mean Gaussian distribution defined in Eq. \eqref{eq:GaussianNoise}, $t_k$ will have a Gaussian distribution
\begin{equation}
    t_k \sim \mathcal{N}\left(0, \tilde{\mathbf{p}}_k \mathbf{W}^{-1} \tilde{\mathbf{p}}_k^T\right) \,.
\end{equation}
Since a Gaussian distribution is uniquely defined by its mean and variance, the following equation will hold:
\begin{equation}
    \mathbf{g}_k \mathbf{S}^{(k)}\mathbf{W}^{-1}\mathbf{S}^{(k)^T} \mathbf{g}_k^T + \sigma_k^2 = \tilde{\mathbf{p}}_k \mathbf{W}^{-1} \tilde{\mathbf{p}}_k^T \,.
\end{equation}
Therefore,
\begin{equation}
    t_k \sim \mathcal{N}\left(0,\mathbf{g}_k \mathbf{S}^{(k)}\mathbf{W}^{-1}\mathbf{S}^{(k)^T} \mathbf{g}_k^T + \sigma_k^2 \right) \,.
\end{equation}

\section{BONFERRONI CORRECTION}\label{app:Bonferroni}
The hypotheses with Bonferroni correction \citep{bonferroni1936teoria} in Eq. \eqref{eq:correct_hypo} have the following relationship with the original hypotheses in Eq. \eqref{eq:origin_hypo}:
\begin{equation}
\begin{aligned}
    H_0 {=}&  \bigcap\limits_{i=1}^{n}H_0^{(k)} \\
    H_1 {=}&  \bigcup\limits_{i=1}^{n}H_1^{(k)} \,.
\end{aligned}    
\end{equation}
Assume that the probability of type I error of the corrected hypothesis test is $\alpha^*$. Then,
\begin{equation}
    \begin{aligned}
        1-\alpha^* {=}& P(\text{All tests accept}|H_0)  \\
                  {=}& 1-P(\text{At least one test is rejected}|H_0) \\
                  {\geq}& 1-\sum\limits_{i=1}^{n} P(\text{Origin test}~ i ~ \text{is rejected}|H_0) \\
                  {=}& 1-\sum\limits_{i=1}^{n} P(\text{Origin test}~ i ~ \text{is rejected}|H_0^{(k)}) \\
                  {=}& 1-n\alpha \,.
    \end{aligned}
\end{equation}
In addition,
\begin{equation}
    \begin{aligned}
        \alpha^* {=}& P(\text{At least one test is rejected}|H_0) \\
                 {\geq}& P(\text{Origin test}~ i ~ \text{is rejected}|H_0^{(k)}) \\
                 {=}& \alpha \,.
    \end{aligned}
\end{equation}
Therefore,
\begin{equation}
    \alpha \leq \alpha^* \leq n\alpha \,.
    \label{eq:alpha* range}
\end{equation}
To keep the type I error $\alpha^*$ not exceeding $\tau$ (e.g., 0.05),
\begin{equation}
    n\alpha=\tau \,.
\end{equation}
Thus, the type I error of the individual test would be $\alpha=\frac{\tau}{n}$.

\section{Relating jackknife residual and solution separation}\label{app:relation}
The normal equations of the WLS problem for the full set and $k$th subset measurements are given by 
\begin{subequations}
    \begin{align}
        \mathbf{G}^T \mathbf{W} \mathbf{G} \hat{\mathbf{x}} &= \mathbf{G}^T \mathbf{W} \mathbf{y} \\
        \mathbf{G}^T \mathbf{W}^{(k)} \mathbf{G} \hat{\mathbf{x}}^{(k)} &= \mathbf{G}^T \mathbf{W}^{(k)} \mathbf{y} \,.
    \end{align}
\end{subequations}
Notably, $\mathbf{W}^{(k)}$ can be re-written by
\begin{equation}
    \mathbf{W}^{(k)} = \mathbf{W} - \mathbf{W}^{\frac{1}{2}}\mathbf{e}_k \mathbf{e}_k^T \mathbf{W}^{\frac{1}{2}},
    \label{eq:W_relation}
\end{equation}
where $\mathbf{W}^{\frac{1}{2}}$ is the square root matrix of $\mathbf{W}$ and $\mathbf{e}_k$ as the $k$th column of the identity matrix. By subtracting the subset normal equation from the full normal equation, we have
\begin{equation}
    \mathbf{G}^T \mathbf{W}\mathbf{G} \hat{\mathbf{x}} - \mathbf{G}^T \mathbf{W}^{(k)} \mathbf{G} \hat{\mathbf{x}}^{(k)} = \mathbf{G}^T \mathbf{W} \mathbf{y} - \mathbf{G}^T \mathbf{W}^{(k)} \mathbf{y} \,.
    \label{eq:normal_eq_substract}
\end{equation}
By substituting Eq. \eqref{eq:W_relation} into Eq. \eqref{eq:normal_eq_substract}, the right-hand side of Eq. \eqref{eq:normal_eq_substract} can be simplified as follows:
\begin{equation}
\begin{aligned}
    \mathbf{G}^T(\mathbf{W}-\mathbf{W}^{(k)}) \mathbf{y}  &= \mathbf{G}^T\mathbf{W}^{\frac{1}{2}}\mathbf{e}_k \mathbf{e}_k^T \mathbf{W}^{\frac{1}{2}} \mathbf{y}  = (\mathbf{W}^{\frac{1}{2}}\mathbf{G})^T\mathbf{e}_k\cdot \mathbf{e}_k^T\mathbf{W}^{\frac{1}{2}}\mathbf{y} \\
    &= W_{k,k}^{\frac{1}{2}}\mathbf{g}_k^T\cdot W_{k,k}^{\frac{1}{2}}y_k=W_{k,k}\mathbf{g}_k^Ty_k \,.
\end{aligned}
\end{equation}
Similarly, the left-hand side of Eq. \eqref{eq:normal_eq_substract} can be simplified as follows:
\begin{equation}
\begin{aligned}
    \mathbf{G}^T \mathbf{W}\mathbf{G} \hat{\mathbf{x}} - \mathbf{G}^T (\mathbf{W} - \mathbf{W}^{\frac{1}{2}}\mathbf{e}_k \mathbf{e}_k^T \mathbf{W}^{\frac{1}{2}}) \mathbf{G} \hat{\mathbf{x}}^{(k)} &= \mathbf{G}^T \mathbf{W}\mathbf{G} (\hat{\mathbf{x}}-\hat{\mathbf{x}}^{(k)}) + \mathbf{G}^T \mathbf{W}^{\frac{1}{2}}\mathbf{e}_k \mathbf{e}_k^T \mathbf{W}^{\frac{1}{2}} \mathbf{G} \hat{\mathbf{x}}^{(k)} \\
    &=\mathbf{G}^T \mathbf{W}\mathbf{G} \mathbf{d}_k +  (\mathbf{W}^{\frac{1}{2}}\mathbf{G})^T\mathbf{e}_k\cdot \mathbf{e}_k^T(\mathbf{W}^{\frac{1}{2}}\mathbf{G}) \hat{\mathbf{x}}^{(k)} \\
    &=\mathbf{G}^T \mathbf{W}\mathbf{G} \mathbf{d}_k+W_{k,k}\mathbf{g}_k^T \mathbf{g}_k \hat{\mathbf{x}}^{(k)} \,.
\end{aligned}
\end{equation}
Therefore, Eq. \eqref{eq:normal_eq_substract} can be written by
\begin{equation}
    \mathbf{G}^T \mathbf{W}\mathbf{G} \mathbf{d}_k+W_{k,k}\mathbf{g}_k^T \mathbf{g}_k \hat{\mathbf{x}}^{(k)} = W_{k,k}\mathbf{g}_k^Ty_k \,.
    \label{eq:sim_normal_eq_substract}
\end{equation}
By substituting Eqs. \eqref{eq:prediction} and \eqref{eq:JK_residual} into Eq. \eqref{eq:sim_normal_eq_substract} , we have
\begin{equation}
    \mathbf{G}^T \mathbf{W}\mathbf{G} \mathbf{d}_k =\mathbf{g}_k^T W_{k,k} t_k \,.
\end{equation}
Since $\mathbf{G}^T \mathbf{W}\mathbf{G}$ is invertible, we have
\begin{equation}
    \mathbf{d}_k = (\mathbf{G}^T \mathbf{W}\mathbf{G})^{-1}\mathbf{g}_k^T W_{k,k} t_k \,.
\end{equation}

\newpage
\section{RESULTS OF OVERBOUNDING CORRECTED PSEUDORANGE ERRORS AT CHTI STATION}\label{app:CHTI_bounding}
\begin{figure}[!htb]
\centering
\includegraphics[width=\textwidth]{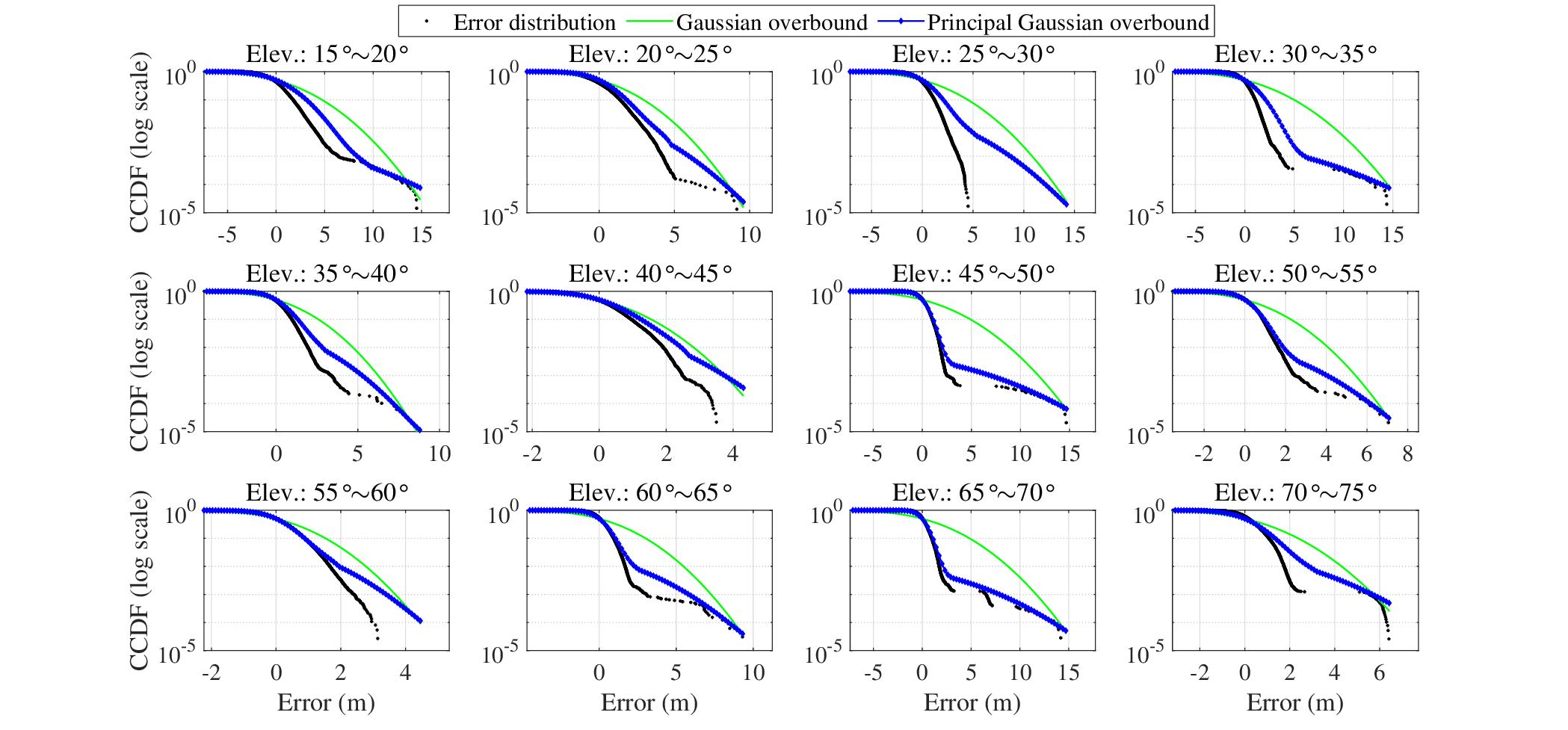}
\caption{The complementary cumulative distribution function (CCDF) of the Gaussian overbound and Principal Gaussian overbound for corrected pseudorange errors in each elevation angle bin. The CCDF is plotted in logarithm scale.}
\label{fig:CHTI_CCDF_compare}
\end{figure}

\end{document}